\documentclass[runningheads]{llncs}

\usepackage[T1]{fontenc}

\usepackage{graphicx}

\usepackage[hyphens]{url}
\begin{document}

\title{Clustering Without Knowing How To: Application and Evaluation}

\author{Daniil Likhobaba\orcidID{0000-0002-0322-3774} \and
Daniil Fedulov\orcidID{0000-0001-7177-8619} \and
Dmitry Ustalov\orcidID{0000-0002-9979-2188}}

\authorrunning{D. Likhobaba et al.}

\institute{Toloka\\\email{\{likhobaba-dp,mr-fedulow,dustalov\}@toloka.ai}}

\maketitle

\begin{abstract}
Clustering plays a crucial role in data mining, allowing convenient exploration of datasets and new dataset bootstrapping. However, it requires knowing the distances between objects, which are not always obtainable due to the formalization complexity or criteria subjectivity. Such problems are more understandable to people, and therefore human judgements may be useful for this purpose. In this paper, we demonstrate a scalable crowdsourced system for image clustering, release its code at \url{https://github.com/Toloka/crowdclustering} under a permissive license, and also publish demo in an interactive Python notebook. Our experiments on two different image datasets, dresses from Zalando's FEIDEGGER and shoes from the Toloka Shoes Dataset, confirm that one can yield meaningful clusters with no machine learning purely with crowdsourcing. In addition, these two cases show the usefulness of such an approach for domain-specific clustering process in fashion recommendation systems or e-commerce.

\end{abstract}

\section{Introduction}

Clustering is a task of grouping objects in such a way that objects in the same group (called a \emph{cluster}) are more similar to each other than to those in other groups \cite{Rokach2005}. This is important process in machine learning and arises in many applications, such as text \cite{text_segmentation} and image \cite{Image_segmentation_by_clustering} segmentation, information retrieval \cite{Content_Based_Image_Retrieval_by_Clustering}, data mining \cite{data_mining} and pattern recognition \cite{importance_of_clustering}. In most cases, clustering is an unsupervised task \cite{unsupervised_clustering} that requires knowing the distances between objects \cite{clustering_distance}.
As the distances are often unknown or the clustering rules cannot be clearly defined \cite{Survey_on_clustering_methods}, crowdsourcing may help to cope with these problems as such tasks often are trivial for humans \cite{A_Survey_of_Crowdsourcing_Systems}. It is known that people can apply their life experience to solve creative tasks \cite{Crowdsourcing_Collaboration_and_Creativity}, such as toxicity detection \cite{Crowdsourcing_Subjective_Tasks_Toxicity_Understanding}, relative rankings \cite{Rankr_Crowdsourcing_opinions}, fashion recommendation \cite{Crowdsourcing_Subjective_Fashion}, etc.

The first idea for obtaining human judgements is to request labels from multiple humans, which allows computational quality control \cite{Ustalov:21:crowdkit}. However, this approach is unscalable, unlike crowdsourcing. Although a proper use of crowdsourcing requires a careful task design and quality control setup, recent studies show that it can approximate the distance function between the objects using crowd judgments \cite{Semi_crowdsourced_clustering}\cite{Crowdclustering_with_partition_labels}\cite{Alloy_Clustering_with_crowds_and_computation}. Some of these papers are theoretical \cite{Clustering_with_Noisy_Queries}\cite{Budget_optimal_clustering_via_crowdsourcing}, evaluate performance on synthetic datasets \cite{Crowdsourced_Clustering_Querying_Edges_vs_Triangles}, or require a prohibitively large number of human tasks to converge \cite{Clustering_with_a_Faulty_Oracle}.

In this demonstration paper, we build a system for clustering with crowds, and evaluate it with crowds without involving any machine learning algorithms. We run our experiments on Toloka with two real world datasets, dresses from Zalando's FEIDEGGER \cite{Feidegger} and shoes from the Toloka Shoes Dataset \cite{Toloka_tutorial}, and confirm the reproducibility of this method. Also, we release the source code of the built hybrid human-computer system under a free license and publish demo in an interactive Python notebook on Google Colab. We picked the \emph{clustering by style} task as it is difficult to formalize as an algorithm, yet the task itself is relatively easy for humans: each of us can tell whether the style of clothes is similar or not.

\section{Task Design and Worker Selection}

To cluster objects, it is necessary to know how similar they are to each other; in the classical formulation, the pairwise matrix of distances is given. If the matrix is not known, it can trivially be approximated with crowds by running a pairwise comparison of all object pairs \cite{Clustering_with_a_Faulty_Oracle}. Unfortunately, it generates $\mathcal{O}(N^2)$ tasks for $N$ objects, which is very expensive, i.e., 1000 objects would require roughly 500,000 comparisons, quickly impacting the annotation budget. Hence, there is a need for a task sampling method that is sufficient to divide the objects into meaningful groups as cheaply as possible.

\subsection{Object Sampling} For clustering, we used Crowdclustering approach \cite{Crowdclustering}. For each task, we show $M$ objects and ask the crowd workers to combine them into groups based on the similarity of style. During prototyping, we found that the optimal choice of $M$ is between 3 and 8 as clustering a large number of objects seems to require additional concentration from the workers, resulting in mistakes, such as failing to group all similar objects. In our setup, every task is completed by three different workers. We sampled each object for $V = \log_2 N \times \log_M N$ times to gather enough information on inter- and intra-relationships of the groups, allowing us to approximate the clustering.

\subsection{Worker Training} Before starting, workers have to pass a training and a qualification test. The training consists of five pages of tasks, each have more pictures and requires more complex actions than the previous one. Starting with two images on the page and a step-by-step guide and ending with six pictures with more complex instructions.
We made instruction and training tasks to familiarise workers with the interface and give them our understanding of how to group clothes by style, while leaving room for their subjective opinion. In the training and exam, workers receive a numerical skill value from 0 to 100, equal to their fraction of correct responses. Only those who achieved the skill value of at least 80 get access to the next step. Training tasks are obvious to everyone, so attentive workers do everything right, and we filter out those who did not understand the task at all. An example of an obvious task is ``label all the high-heeled shoes with red color from palette.''

\begin{figure}[h]
\centering
\includegraphics[width=\columnwidth]{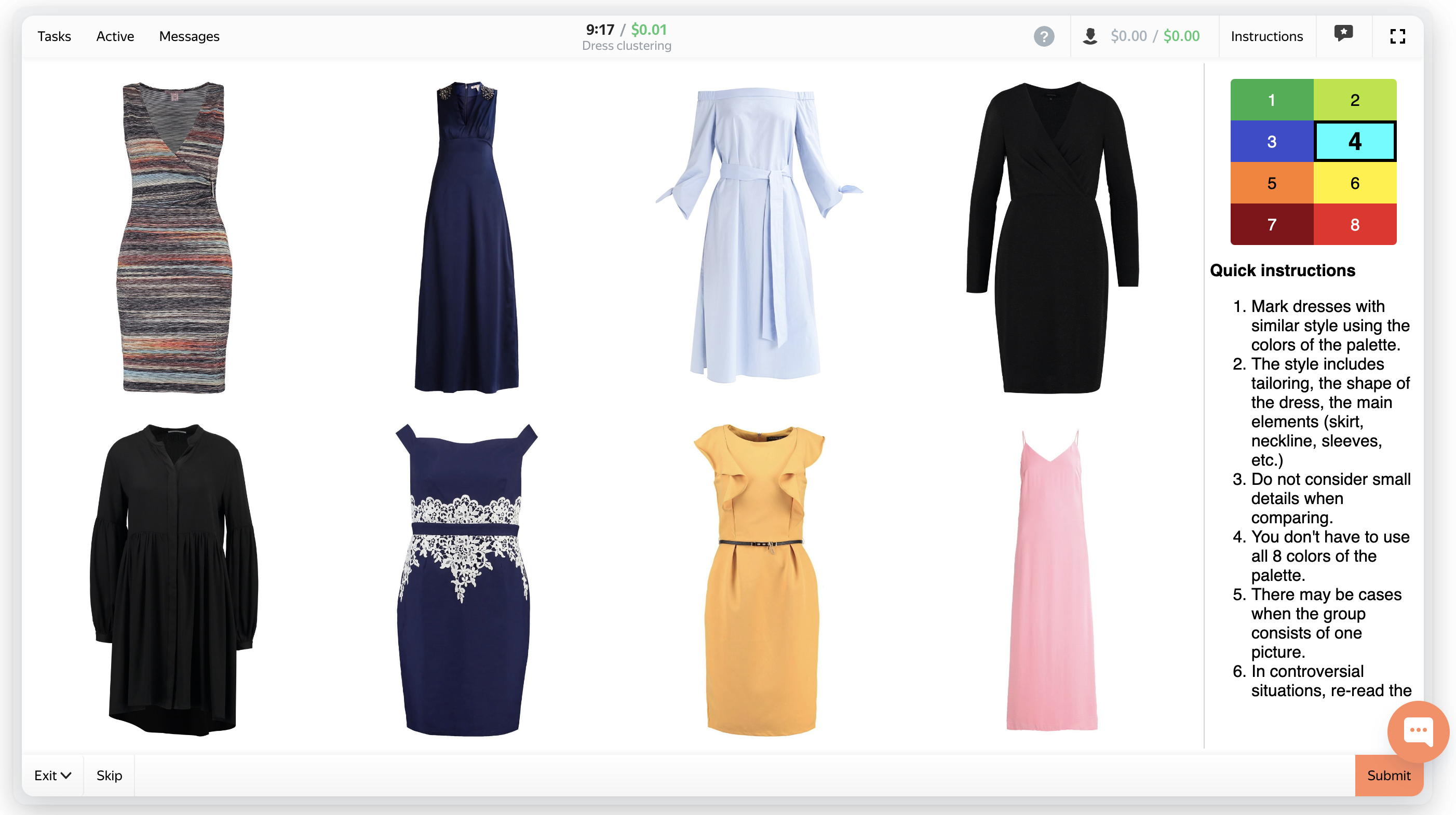}
\caption{\label{task_interface}Clustering Task. The worker groups similar clothes by highlighting similar ones with same color from the color palette; similar to \cite{Crowdclustering}.}
\end{figure}

\subsection{Task Design} The task is formulated as \emph{Group the objects by labeling similar ones using color palette}, and its interface is shown in Figure~\ref{task_interface}. Workers should choose one color and label similar images with it, then choose another color and make another group, etc. This color palette simply serves as labels that are convenient for a worker, and \emph{it has nothing in common with the colors of objects}. We then treat these colors simply as numbers assigned to each object on the page, placing objects with the same numbers in the same list for each page. Since each item is completely unique, the workers are told not to pay attention to the small details when grouping clothes, but to look at the style as a whole. There is a brief instruction on each page with the main points that should be kept in mind during grouping.

\section{Clustering with Crowds}

After labeling, we have information about how the pictures on each page have been grouped together, and which worker made each page. Since we have a sparse dataset of noisy labels for the objects, we need a special aggregation method to recover the clusters. For this, we applied and re-implemented using Python a probabilistic model called \emph{Crowdclustering} \cite{Crowdclustering}. This approach represents the objects as points in Euclidean space and models the workers as a plurality of binary classifiers. Each task is considered as a binary classification of the objects pairs regarding their membership in the same group or in a different group.
The main advantage of Crowdclustering algorithm is that it allows each worker to group objects by any attribute (e.g. color, material, shape) and works with these groups called \emph{atomic clusters}. Then, atomic clusters are assembled into resulting clusters, the number of which is not a fixed hyper-parameter and is estimated dynamically.

\begin{figure}[h]
\centering
\includegraphics[width=\columnwidth]{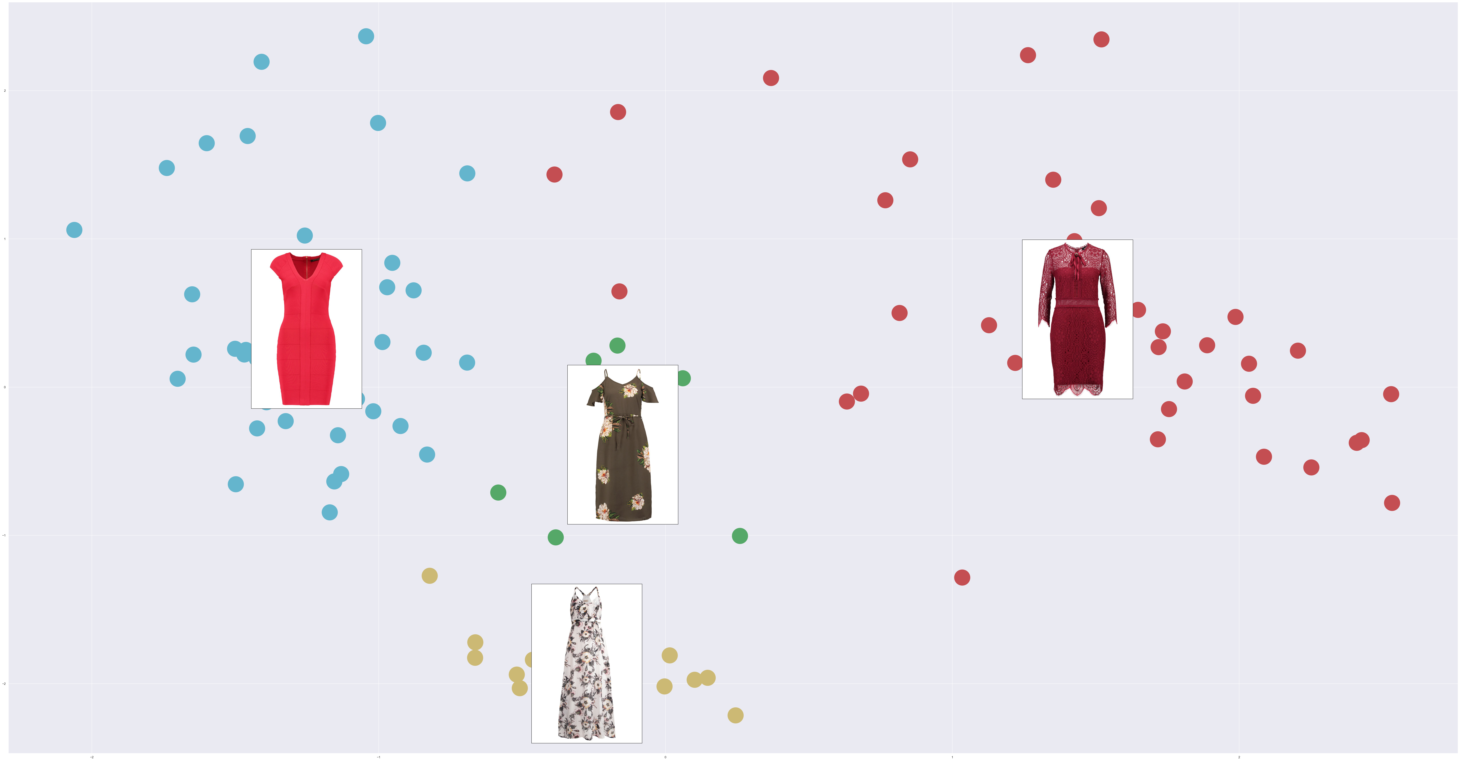}
\caption{\label{result_clusters}Cluster visualization produced by the Crowdclustering method; dots are objects, colors are clusters.}
\end{figure}

\section{Quality Evaluation}

\begin{figure}[t]
\centering
\includegraphics[width=\columnwidth]{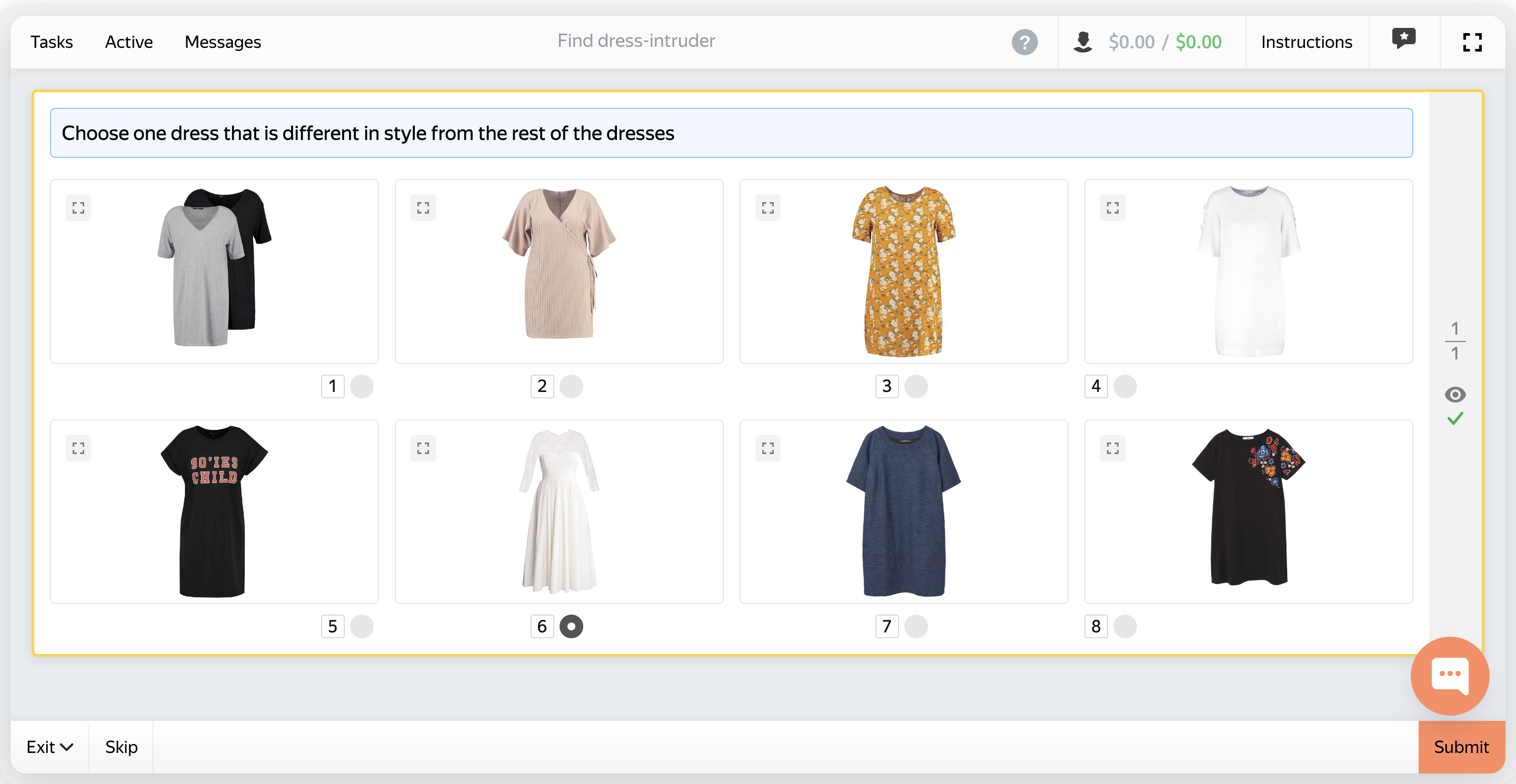}
\caption{\label{dress_intruder}Evaluation Task. The worker has to select the artificially-inserted intruder element (no. 6); similar to \cite{Intruders}.}
\end{figure}

Having annotated and aggregated the clusters, we need to evaluate the quality of them. A common approach is to compare result with ground truth answers. However, we are solving a problem without a strict criterion for grouping objects that is difficult to formalise, i.e. \emph{clustering clothes by style}. It leads to difficulties with making the only correct ground truth answers. And since we use subjective human judgements for clustering, we want to use them for evaluation as well. For this purpose we apply an approach called \emph{Intruders} \cite{Intruders}. For each cluster, we sample from another cluster a random incorrect object called an intruder. Then, we run another crowdsourcing task, in which we ask the workers to select the out-of-style object. This task's interface is shown in Figure~\ref{dress_intruder}. The clustering quality is a fraction of times the workers selected the intruder correctly and it is considered the better, the more often the workers choose this obviously incorrect object.

We ran the experiments on FEIDEGGER and Toloka Shoes in the same above-described configuration. Visualization of result is shown in Figure~\ref{result_clusters}, where dots in Euclidean space are projected on a 2D plane for clarity. We found that for the 2,000 dress images in FEIDEGGER the quality is 0.83, and for the 87 shoes images in Toloka Shoes Dataset the quality is 0.88. It means that in the clustering tasks crowd workers divided images into clusters according to their subjective opinion, which are consistent with the subjective opinion of other workers who evaluated these clusters.

\section{Conclusion}

We found that crowdsourcing allows to obtain a reasonable clustering of objects when distances between the objects could not be measured at all due to informal criteria. It allows using a human-understandable textual instruction instead of metric learning, while being more cost-efficient than the entire distance matrix annotation. This may be important when the problem is difficult to formalize for machine learning algorithms. We release a Python implementation of the system at \url{https://github.com/Toloka/crowdclustering}.
It includes the labeling pipeline, a Crowdclustering algorithm, and quality evaluation algorithm.

\bibliographystyle{splncs04}
\bibliography{clustering}

\end{document}